THEME ARTICLE: HUMAN MEETS AI

# Assessing Latency in ASR Systems: A Methodological Perspective for Real-Time Use

Carlos Arriaga, Alejandro Pozo, Javier Conde, and Alvaro Alonso, *Universidad Politécnica de Madrid, 28040, Madrid, Spain*

*Automatic speech recognition (ASR) systems generate real-time transcriptions but often miss nuances that human interpreters capture. While ASR is useful in many contexts, interpreters—who already use ASR tools such as Dragon—add critical value, especially in sensitive settings such as diplomatic meetings where subtle language is key. Human interpreters not only perceive these nuances but can adjust in real time, improving accuracy, while ASR handles basic transcription tasks. However, ASR systems introduce a delay that does not align with real-time interpretation needs. The user-perceived latency of ASR systems differs from that of interpretation because it measures the time between speech and transcription delivery. To address this, we propose a new approach to measuring delay in ASR systems and validate if they are usable in live interpretation scenarios.*

Automatic speech recognition (ASR), also known as speech-to-text (STT), uses artificial intelligence to convert human speech into text. ASR models have evolved from pipeline-based architectures[1] to end-to-end (E2E) deep learning models,[2] such as long short-term memory, recurrent neural network transducer (RNN-T), and Transformers.[3]

STT interpretation is the generation of a written version of a speech in real time. Before the development of ASR, STT interpretation had to be performed by professional interpreters with an adapted keyboard. The development of automatic STT systems can facilitate the work of STT interpreters by providing preliminary transcriptions that interpreters can edit. In addition, STT systems can generate automatic transcriptions, making the access to audio-based Web applications, such as online education[4] and virtual meetings, more inclusive for people who are deaf or hard of hearing, such as the elderly.[5]

However, both of these applications have strict E2E delay requirements. End users need the transcriptions to be generated during the speech to allow them to participate in the conversation, so interpreters' STT tools have to transcribe during the speech. If the ASR systems have a high latency, the collaboration between ASR tools and STT interpreters is not possible. The incoming speech must be fragmented into smaller segments that are sequentially processed. The speech cannot be separated at any point, as splitting a word into two fragments results in an incorrect transcription. The algorithm must minimize the time the audio is buffered and maximize the quality of the transcription.

In this article, we review the existing models and metrics used for real-time ASR and study their usage in STT interpretation. After that, we propose a methodology to evaluate these systems and determine their viability in real-time STT interpretation.

## RELATED WORK AND PROPOSAL

STT translation has strict E2E delay restrictions, as transcriptions must be generated as the user speaks to enable live interaction. In traditional interpretation, a commonly used metric to evaluate the delay is the ear-voice-span (EVS), which refers to the time lag between an interpreter hears a fragment and when they begin to translate it. However, most metrics used to evaluate ASR systems are not applicable to real-time systems, and there is not an equivalent measure to the EVS. A common metric used is the real-time factor (RTF), which is defined as the time of an utterance divided by the time it takes the system to process it. A system is considered real-time ready if the RTF is less than one, which means that the system can transcribe a speech faster than it is being spoken. However, this metric does not take into account other factors that affect the E2E delay, such as the time the samples are buffered before being processed.

To reduce the latency of ASR systems, researchers have analyzed improving the endpointer of the model[6] and adapting E2E models to streaming scenarios by applying time restrictions and late generation penalties.[7] However, this research mainly focuses on reducing the delay between the end-of-speech (EOS) and the generation of the transcription. None of these articles provide an E2E delay that takes into account the times audio samples are buffered and transmissions delay. For these reasons, in this article, we propose a new definition of E2E delay for real-time ASR systems.

For these reasons, the proposed measure is based on the user-perceived latency (UPL) and aims to be the equivalent to the EVS in traditional interpretation. The EVS cannot be used directly as a delay measure in ASR systems. The EVS takes into account factors such as the time interpreters retain the information until they start to translate and the time it takes to process it. However, as stated before, it does not take into account the transmission delay or the time the samples are buffered before being sent to the model.

The UPL of an ASR system varies depending on the applications. For example, for command-based systems, the UPL is the time elapsed between the EOS and the transcription is finished.[8] The UPL in ASR systems must be the delay between the speaker uttering a word and the end user receiving it.

The delay is not the only metric that determines the viability of an ASR system. The quality of the generated transcription must also be considered. Regarding real-time ASR models, the focus is on architectures such as connectionist temporal classification or RNN-T. However, transformer-based ASR systems have shown great results for offline ASR in conjunction with ease of training. Over the last years, transformer-based models such as Whisper[9] have obtained great results in terms of transcription quality with platforms such as Whisper-streaming,[10] Simul-Whisper,[11] or Whisper-flow.[a]

In this article, we focus on providing a new definition of delay in real-time ASR systems. First, the architecture of these ASR systems is analyzed to determine the sources of the delay. After that, a new definition of delay is provided. The proposed methodology is then tested with three different algorithms to validate the methodology and establish the basis for the evaluation of ASR systems.

[a] https://github.com/dimastatz/whisper-flow

## METHODOLOGY DEFINITION

The methodology measures the performance of the models for STT interpretation, where the audio has to be transcribed during the speech. However, the E2E delay is only one of the parameters used to determine the viability of an ASR system. The parameters measured to determine the quality of transmission are as follows:

- *Model performance*: To measure the performance of the models, three different parameters are measured: word error rate (WER), match error rate (MER), and word information loss (WIL).[12] These metrics are based on the Levenshtein distance and require aligning the resulting transcription with the ground truth.
- *E2E delay*: This is the delay since a word is uttered until the transcription appears in the client application.
- Quality/delay: For real-time scenarios, the relationship between transcription accuracy and delay is studied.

The goal is to analyze the relationship between delay and quality, as it has to provide both a low enough delay and sufficient quality to be usable in real-time STT.

### ASR Architecture

Multiples systems such as Whisper-streaming, Simul-Whisper, or Whisper-flow have been developed to adapt Whisper to real-time scenarios. Since Whisper is designed for offline translation, all these implementations follow the same architecture:

- *Audio input:* The audio that enters the system to be transcribed.
- *Audio extractor*: The process that captures the audio from the audio input and transforms it into a suitable format for the rest of the system.
- *Audio splitter*: The process that receives the audio from the audio extractor and optimally splits it into fragments for processing.
- *ASR cluster*: Responsible for transcribing the received audio samples.
- *ASR cluster-audio splitter connection*: The connection between the ASR cluster and the audio splitter, used to transmit the audio fragments.

- *Display*: The component that receives the transcription and renders it.
- *ASR cluster-display connection*: The connection between the ASR cluster and the display, used to transmit the resulting transcription.

## E2E Delay Definition

We define delay in a real-time system transcription system as the time elapsed between a user pronounces a word and its transcription is presented to the user. This means that systems that release output in large bursts, despite low internal processing delay, will be evaluated as having higher UPL due to the perceptual delay introduced. For STT interpretation, average lagging has been selected over other measures such as length-adaptive average lagging,[13] streamingLAAL[14] so it is not expected to over-generate or under-generate. The total delay ($D_T$) is defined as

$$D_T = D_s + D_p + D_t. \quad (1)$$

The definition of each term of the equation is as follows:

- The splitting delay ($D_s$) is the delay caused by the audio splitter. Audio samples are stored until the algorithm generates a fragment and sends it to processing.
- Processing delay ($D_p$) accounts for the delay caused by the audio processing. It includes the time taken for the audio extractor to convert the input and the time taken by the ASR cluster to generate the transcription.
- Transmission delay ($D_t$) is the lag introduced by the splitter-ASR cluster connection and the ASR cluster-display connection.

Each value of the equation depends on a different factor. $D_s$ is independent of hardware, since it depends only on the time the audio splitting algorithms buffer the samples. $D_p$ varies depending on the hardware used to run the models and the different processes. $D_t$ depends on the geographical position of the components and the connection between them. It is independent of both algorithm and model.

This measure is an adaptation of the EVS metric that accounts for the additional delays introduced by these digital systems. In traditional on-site interpretation, $D_t$ can be ignored as interpreters, speakers, and the public are co-located at the same place. In addition, the audio splitter does not exist in traditional interpretation, as interpreters are constantly receiving the speaker's audio, so $D_s$ can also be ignored.

## Delay Measurement Algorithm

To accurately measure the transcription delay, reference transcriptions must include time annotations.

Every time a transcription is rendered on the display, all of the words included in this fragment must be stored with their timestamp. Words that appear at the start of a fragment will have an increased delay as their audio samples have to be stored until the fragment is complete. For this reason, the process must be repeated multiple times to average the results.

Because in the resulting transcription, each word does not only appear once (words such as "the" or "it" have a high number of repetitions), the correct repetition must be found to compare timestamps. For this reason, words must be searched with their context.

We define context as the $M$ words before and after the selected word. To reduce false positives due to common expressions such as "I am not," "there were," and so on, a sliding-window algorithm can be used to reduce the search width.

For the first word searched, the algorithm checks if it exists in the first $N$ segments. All of the segments where the word is found are stored. After that, for each of those segments, we search for the word with its context. For the next iteration, the next word is searched in the range $[i, N+i]$, with $i$ being the index of the last word found in its context, and $N$ the width of the search. In the case where a word is not found, $N$ is increased until a new match is found.

## ASR Models

The selected ASR for the model will affect both the resulting quality and the delay of the system. Larger models usually have a better performance at the cost of requiring longer processing times. So, for a selected audio splitting algorithm, a larger model is expected to perform better. This hypothesis will be tested in the "Methodology Application" section.

## Testing Process

After analyzing the ASR architectures, the causes of the delays, and the impacts of both the audio-splitting algorithms and the ASR model, the methodology can be defined.

In the previous sections, we have identified four variables that affect the delay or the quality of the system:

- *Hardware*: Affects $D_p$.
- *Audio splitting algorithm*: Affects both $D_s$ and the resulting quality.
- *Transmission delay*: Affects $D_t$.

- *ASR model*: Affects both $D_p$ and the resulting quality.

Of these four variables, the transmission delay is difficult to evaluate in a local environment, as it depends on the location of the user. Because of this, developers must consider the worst-case scenario for their use case and add this value to the delay obtained in the local tests.

Because of the hardware dependency, the following steps must be repeated for each hardware setup. First, every ASR model must first be tested in an offline scenario. Models with an RTF greater than one must be discarded as they are not able to generate transcription fast enough for real-time interpretation. After that, the quality of the transcriptions must be measured. Models that do not provide sufficient WER must be discarded, as their performance in online interpretation will be degraded.

Finally, because of the interconnection between the ASR model and the audio splitting algorithm, every combination of the model–algorithm must be independently tested. Both the E2E delay and the transcription quality must be measured as proposed to determine the viability of the combination. The minimum WER and maximum delay depends on the application and its intended usage, so no values are provided in this methodology.

### Algorithms and Model Comparison

We define that an algorithm–model combination $C_1$ is better than another combination $C_2$ only if it has lower delay and lower WER.

For example, in a professional environment, combinations with lower WER can be better, as the quality of the transcription must comply with a certain quality standard despite the delay.

## METHODOLOGY APPLICATION

To validate our approach, we performed an experiment with three Whisper models using the proposed methodology. We selected the tiny,[b] base,[c] and large[d] models from OpenAI, to compare models of different sizes.

The dataset used for these experiments was GigaSpeech.[15] This repository includes 10,000 h of transcribed audio. We selected the 52 h from the testing and evaluation subset as they are annotated by professional humans and are stated to provide better coverage on both topics and audio conditions. The dataset includes audio from podcasts and YouTube videos, some of them with background noise, to resemble real-life situations. GigaSpeech transcriptions are divided into segments with start and end timestamps. For this reason, the only known timestamps are when the first word of the segment starts to be pronounced and when the last word is pronounced. Because of this, to measure the delay, we only use the first word for each segment in the delay measurement.

These experiments can be repeated using other datasets that include word-level timestamps or by using force alignment to generate the timestamps.

Gigaspeech's transcriptions and the ones generated by Whisper have different formats, so we performed a normalization process: Numbers were replaced by their textual version, annotations were removed, punctuation symbols were standardized, and so on.

All components defined in the architecture were deployed on an Intel Core i9-12900 with 3x8-GB RAM M323R1GB4BB0. For this reason, the specific delays measured are only applicable for this hardware, even though comparisons between the different combinations of models-algorithms can be performed. As stated in the methodology, $D_t$ is omitted and must be taken into account in a real scenario.

### Audio Splitting Algorithms

In this section, the three tested algorithms are presented. The goal of these models is to illustrate the delay introduced by the audio splitter and the effects of the selected algorithm in the quality of the transcription.

#### *Fixed Interval*

A first naive algorithm splits audio fragments at fixed intervals. Once the interval ends, all stored samples are sent for transcription. Different intervals will be tested to measure the impact of the duration of the fragment on delay and performance. Two different fragmentation intervals, 2 and 3 s, were tested to compare its effects on the resulting transcription.

#### *VAD Based*

The voice activity detection (VAD)-based algorithm tries to solve the word-splitting issues caused by the previous algorithm. In contrast, with end-of-query detectors, VAD algorithms detect the silence between word instead of only detecting the end of the speech. In this implementation, a JavaScript VAD library[e] is used.

[b] https://huggingface.co/openai/whisper-tiny
[c] https://huggingface.co/openai/whisper-base
[d] https://huggingface.co/openai/whisper-large-v2
[e] https://github.com/kdavis-mozilla/vad.js

The VAD is a state machine with two states: silence and voice. The samples are stored in both states. When the transition from voice to silence is triggered, all stored samples are sent to the ASR cluster.

### *Feedback*

The last algorithm tries to reduce the delay of the VAD-based algorithm while maintaining quality. Instead of waiting for silence, samples are sent at a fixed interval, but to avoid word separation, some audio samples from the previous fragment are reused. This is an adaptation from the algorithm used by Whisper-streaming where the feedback is of fixed size instead of the complete sentence, as it caused long processing delays for longer sentences.

To merge both transcriptions, we use a LocalAgreement-2 policy, as represented in Figure 1. LocalAgreement has outperformed other popular partial hypothesis selection strategies such as wait-k or hold-n in low-latency scenarios.[16]

For the feedback algorithm, words that were replaced by the merge process are excluded for the delay measurement. Following the example in Figure 1, only the words "speedcuber. The vision for the" were stored with their corresponding timestamp.

## Offline Test

To compare the performance of the various audio splitting algorithms and models, the audio files were transcribed one by one using a shell script. The large model was discarded as the CPU was not able to run the model with an RTF lower than one.

## STT Interpretation Testing

To create an environment that simulates real-world usage, we used Selenium,[f] an open source project to automate browsers. For each audio file in the dataset, a headless browser was launched with the audio file as the microphone. The captured audio was transcribed until the file ends.

# RESULTS

## Model Performance

The WER, WIL, and MER are measured for each model as proposed in Morris et al.[12] The results obtained for all experiments are collected in Table 1. The library that calculates the WER, MER, and WIL is Jiwer.[g]

As observed, the best WER, MER, and WIL are obtained by the large model in batch mode. When comparing transcription quality, larger models provide better results as expected. All models see their performance degraded when executed in real-time scenarios. Depending on the audio splitting algorithm, this increase varies. The best results for both the tiny and base models are obtained with the VAD algorithm, followed by the feedback algorithm. Even with the word correction performed in the feedback algorithm, the WER is 3.57% and 2.32% higher than with VAD splitting.

The fixed-interval algorithm performs the worst of the three, as utterances can be separated into different fragments. Comparing the 2-s fixed splitting to VAD splitting, there is an increase of 9.07% and 10.82% for the tiny and base models, respectively. It can be observed that shorter split intervals generate worse results than larger intervals. Increasing the splitting interval from 2 to 3 s reduces the WER in 4.08% and 6.51% for the tiny and base models, respectively. This is because with larger fragments, less fragments are generated, so fewer utterances can be incorrectly divided.

## E2E Delay

The delays obtained with the three algorithms are presented in Table 1. As stated in the "Methodology Definition" section, These values will vary depending on the hardware, as the processing delay depends on the hardware.

Results show that, using the same audio splitting algorithm, smaller models result in lower delays. This is expected, as smaller models require less computation time. Comparing algorithms, the VAD algorithm introduces the longest delay of all models, as samples are stored until a silence is found.

For fixed audio-splitting measures, the minimum expected delay was half the segment duration. This is the average time since a word is pronounced until the fragment is sent for processing. All four measures taken comply with it. The results show that larger audio fragments produce more accurate transcriptions.

The performance of the feedback algorithm is placed in the middle. Even though audio fragments

Previous: speedcuber. The vision for the ~~walk trough~~

New: ~~mission~~ for the Volk III is brought about

Merge: speedcuber. The vision for the Volk III is brought about are sent at the same rate as in the 2-s fixed algorithm, due to the 4 s of feedback introduced, it takes longer to process the fragments. There is a significant decrease of 1.521 s and 1.987 s in delay in the feedback algorithm compared to the VAD algorithm for the tiny and base mode. As measured in the previous section, this is exchanged for a WER increase of 3.57% and 2.32%. In relative terms, this means that the feedback algorithm takes 43.46% and 44.33% less time than the VAD algorithm for an increase of 13.99% and 10.50% in the WER.

**FIGURE 1.** Merge process of the previous and new transcription using LocalAgreement-2.

[f]https://www.selenium.dev/

[g]https://pypi.org/project/jiwer/

**TABLE 1.** WER, WIL, and MER of the different combinations of model and audio splitting algorithms.

| Model | Audio splitting algorithm | WER | MER | WIL | Delay (ms) | Delay measures |
|---|---|---|---|---|---|---|
| Large | Batch | 0.1323 | 0.1249 | 0.1718 | — | — |
| Tiny | Batch | 0.1748 | 0.1672 | 0.2326 | — | — |
| Tiny | 3-s fragment | 0.3050 | 0.2834 | 0.3928 | 2244 | 679 |
| Tiny | 2-s fragment | 0.3458 | 0.3127 | 0.4390 | 1702 | 369 |
| Tiny | VAD | 0.2551 | 0.2396 | 0.3358 | 3521 | 1066 |
| Tiny | Feedback | 0.2908 | 0.2679 | 0.3516 | 2000 | 367 |
| Base | Batch | 0.1646 | 0.1574 | 0.2126 | — | — |
| Base | 3-s fragment | 0.2735 | 0.2544 | 0.3461 | 2783 | 744 |
| Base | 2-s fragment | 0.3386 | 0.3089 | 0.4136 | 2269 | 391 |
| Base | VAD | 0.2304 | 0.2304 | 0.2971 | 4483 | 1019 |
| Base | Feedback | 0.2536 | 0.2301 | 0.3058 | 2496 | 419 |

The number of measures taken is the number of words that were found in their context. The same 16,985 words were searched for each combination of model–algorithm. Because an exact match of the context was needed, only a small subset of words were found for each combination due to the high WER.

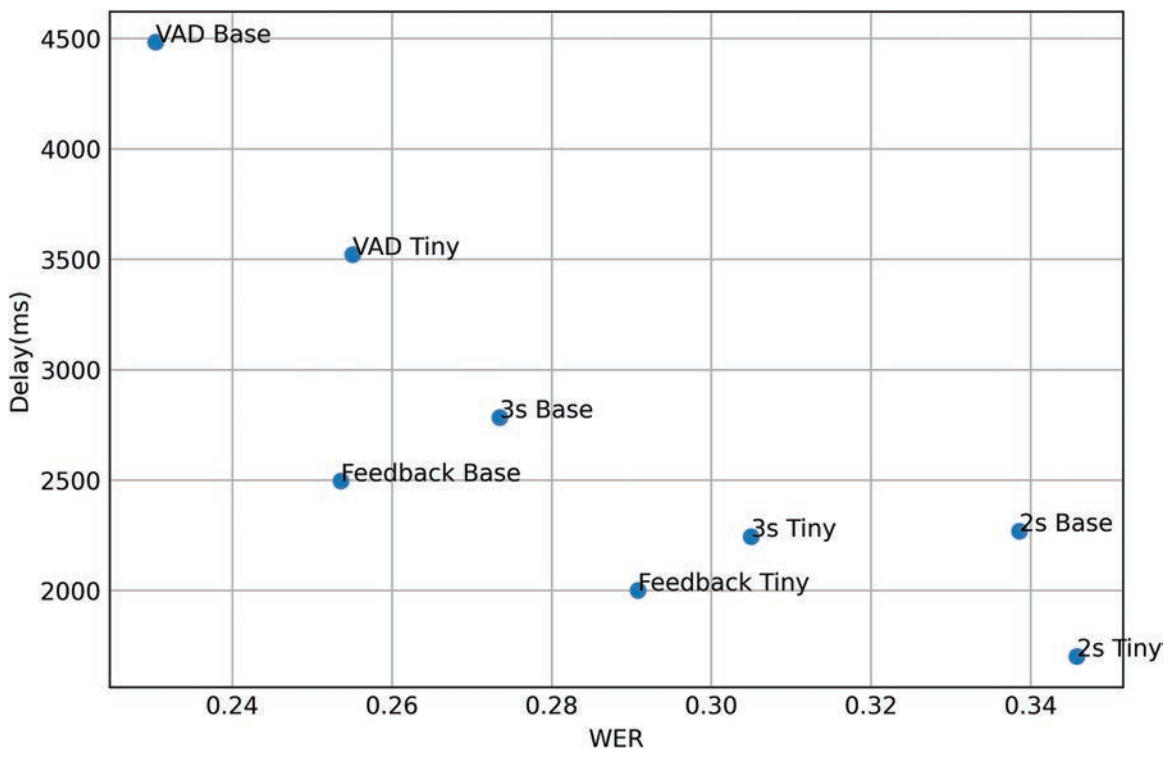

**FIGURE 2.** WER-delay matrix of the different combinations of model and algorithm.

## Quality-Delay Matrix

Finally, to compare the combinations of model–algorithm, following our methodology, we plot the quality and delay of each combination in a 2-D matrix.

All combinations tested are represented in Figure 2. By the definition provided in the "Algorithms and Model Comparison" section, there are only two combinations that can be considered to be better than the other, as having both lower delay and lower WER is required: The combination of 3-s fixed splitting with the tiny model is worse than the tiny model with the feedback algorithm, and the feedback algorithm with the base model outperforms the VAD algorithm with the tiny model and the 3-s fixed splitting with the base model.

With better hardware or GPU acceleration, the resulting delay can be reduced. There is a maximum reduction in delay that can be achieved by improving the hardware as $D_p$ is only one part of $D_T$.

## CONCLUSION

This work presents a new definition of delay tailored for real-time STT systems. The proposed measure adapts the EVS metric by incorporating the additional delays caused by the audio splitter and intercomponent communication. This work serves as a foundation for evaluating ASR systems for real-time STT interpretation and supports the development of more accessible videoconferencing applications and online education tools for hearing-impaired users as well as assisting STT interpreters in their work. Developers can apply this methodology to compare existing models and algorithms, enabling them to select the most suitable combination for their specific use case.

The proposed methodology was used to compare two models with three audio splitting algorithms. In our evaluation, fixed-interval splitting algorithms had the best quality and the highest delay, VAD-based splitting had the worst quality and the lowest delay, and feedback-based algorithm placed in the middle on both metrics.

Future work includes exploring integration with technologies designed for real-time such as WebRTC for videoconferencing and WebAssembly[17] to embed ASR models directly in browsers as well as compatibility with tools commonly used by STT interpreters. Integration with WebAssembly is of particular interest as it can replace the ASR cluster with a local instance of the ASR model. This allows for the system to run without the need of an Internet connection. The performance of this solution must be tested: The potential reduction in delay introduced with the elimination of the ASR cluster–audio splitter connection can be offset by the performance of the model embedded in the browser.

## ACKNOWLEDGMENTS

The authors would like to acknowledge the support of the FUN4DATE (PID2022-136684OB-C22) project funded by the Spanish Agencia Estatal de Investigación (AEI 10.13039/501100011033), the TUCAN6-CM (TEC-2024/COM-460) project funded by CM (ORDEN5696/2024), and the Ministerio de Ciencia, Innovación y Universidades (Grant FPU21/03069).

**CARLOS ARRIAGA** is a Ph.D. candidate at the Universidad Politécnica de Madrid, 28040, Madrid, Spain. His research interests includes videoconferencing and artificial intelligence. Arriaga received his master’s degree in engineering from the Universidad Politécnica de Madrid. Contact him at carlos.arriaga.prieto@upm.es.

**ALEJANDRO POZO** is an assistant professor at the Universidad Politécnica de Madrid, 28040, Madrid, Spain. His research interests include artificial intelligence, data engineering, and cybersecurity. Pozo received his Ph.D. degree in engineering from the Universidad Politécnica de Madrid. Contact him at alejandro.pozo@upm.es.

**JAVIER CONDE** is an assistant professor at the Universidad Politécnica de Madrid, 28040, Madrid, Spain. His research interests include artificial intelligence, open data, and digital twins. Conde received his Ph.D. degree in engineering from the Universidad Politécnica de Madrid. Contact him at javier.conde.diaz@upm.es.

**ALVARO ALONSO** is a professor at Universidad Politécnica de Madrid, 28040, Madrid, Spain. His research interests include multiconferencing systems in cloud computing and artificial intelligence. Alonso received his Ph.D. degree in engineering from the Universidad Politécnica de Madrid. Contact him at alvaro.alonso@upm.es.